\documentclass[journal]{IEEEtran}
\usepackage{algorithm,algorithmic}
\usepackage[moderate]{savetrees}
\usepackage{verbatim}
\usepackage{subcaption}
\usepackage{amsmath,amssymb,amsfonts,bm}
\usepackage{epsf}
\usepackage{tikz}
\usepackage{psfrag}
\usepackage{pstool}
\usepackage{epstopdf}
\usepackage[margin={0.7in}]{geometry}
\usepackage{booktabs}
\usepackage{stfloats}
\usepackage{url}
\usepackage[dvips]{epsfig}
\usepackage{amsthm}
\usepackage{color}
\usepackage[usenames,dvipsnames]{pstricks}
\usepackage{lettrine}
\usepackage[dvips]{epsfig}
\usepackage{pst-grad} % For gradients
\usepackage{pst-plot} % For axes
\usepackage{epsfig}
\usepackage{enumerate}
\usepackage{amsbsy}
\usepackage{amssymb}
\usepackage{amsthm}
\usepackage{amscd}
\usepackage{float}
\usepackage[caption = false]{subfig}
\usepackage{color}
\usepackage[numbers,sort&compress,square]{natbib}

\usepackage{listings}
\usepackage{stackrel}
\usepackage{authblk}
\usepackage{dsfont}
\usepackage[normalem]{ulem}
\makeatletter
\newcommand{\removelatexerror}{\let\@latex@error\@gobble}
\makeatother
\newtheorem{thm}{Theorem}
\newtheorem{lemm}{Lemma}
\newtheorem{cor}{Corollary}

\usepackage{stackengine}

\definecolor{mintbg}{rgb}{.63,.79,.95}
\usepackage{enumitem}
\begin{document}


\title{{Tackling Non-IIDness in HAPS-Aided Federated Learning}}
\vspace{-1mm}
\author{Amin~Farajzadeh,~\IEEEmembership{Member,~IEEE,}
        Animesh~Yadav,~\IEEEmembership{Senior~Member,~IEEE,}
        and~Halim~Yanikomeroglu,~\IEEEmembership{Fellow,~IEEE}% <-this % stops a space
\thanks{A. Farajzadeh and H. Yanikomeroglu are with the Non-Terrestrial Networks (NTN) Lab, Department of Systems and Computer Engineering, Carleton University, Ottawa, ON K1S 5B6, Canada (e-mail: aminfarajzadeh@sce.carleton.ca; halim@sce.carleton.ca). A. Yadav is with the School of EECS, Ohio University, Athens, OH, 45701 USA (e-mail: yadava@ohio.edu).}}% <-this % stops a space
\pagenumbering{gobble}
\makeatletter
\patchcmd{\@maketitle}
  {\addvspace{0.5\baselineskip}\egroup}
  {\addvspace{-1.7\baselineskip}\egroup}
  {}
  {}
\makeatother

\maketitle

\begin{abstract}
High-altitude platform stations (HAPS) enable large-scale federated learning (FL) in non-terrestrial networks (NTN) by providing wide-area coverage and predominantly line-of-sight (LoS) connectivity to many ground users. However, practical deployments face heterogeneous and non-independently and identically distributed (non-IID) client data, which degrades accuracy and slows convergence. We propose a weighted attribute-based client selection strategy that leverages server-side indicators: historical traffic behavior, instantaneous channel quality, computational capability, and prior-round learning contribution. At each round, the HAPS computes a composite score and selects the top clients, while adapting attribute weights online based on their correlation with validation-loss improvement. We further provide theoretical justification that traffic-derived uniformity can serve as a proxy for latent data heterogeneity, enabling selection of client subsets with reduced expected non-IIDness. Simulations demonstrate improved test accuracy, faster convergence, and lower training loss compared with random, resource-only, and single-attribute baselines.
\end{abstract}

\vspace{-1mm}
\vspace{-2mm}

\IEEEpeerreviewmaketitle

\vspace{-2mm}
\section{Introduction}
In the rapidly evolving domain of privacy-preserving machine learning, federated learning (FL) has emerged as a paradigm-shifting approach that develops a global learning model by collaborating among the multitude of clients with unshared data repositories \cite{FL2}. Its performance improves with the number of participating data repositories. To support large-scale FL, non-terrestrial networks (NTN), particularly high-altitude platform stations (HAPS), have been proposed~\cite{NTN_FL_new1}. Operating in the stratosphere, HAPS provide wide-area coverage and reliable line-of-sight (LoS) links to ground devices~\cite{NTN_FL_new2}, significantly expanding the pool of eligible FL clients. This enables more diverse and large-scale participation compared to terrestrial networks, which are constrained by limited coverage and LoS connectivity~\cite{HAPS1}.

Nevertheless, this HAPS-enabled large-scale connectivity inherently amplifies the challenge of non-independently and identically distributed (non-IID) data. Unlike terrestrial settings with relatively localized and homogeneous users, HAPS aggregate geographically dispersed clients with highly diverse traffic patterns, contextual conditions, and sensing environments~\cite{non-IID2}. As a result, client datasets exhibit significant distributional heterogeneity. Incorporating such highly non-IID clients in FL training degrades performance, leading to slower convergence and reduced model accuracy~\cite{non-iid-new}.

To address this challenge, some recent works~\cite{lit1,lit2,lit3,lit5,litNew,lit6} have proposed a variety of client selection strategies for FL. Focusing primarily on optimizing operational constraints like efficiency and reliability for FL, these strategies incorporate methods such as client clustering, reputation, game-theoretic, and stochastic techniques. Among these, cluster FL has emerged as a promising solution for addressing non-IID data by grouping clients with similar data distributions~\cite{lit1},~\cite{lit3}. For instance, \cite{lit6} leverages advanced K-means clustering to group clients based on feature similarity, while \cite{lit3} uses a computationally complex selection method based on probability distributions to reduce convergence biases. These approaches primarily focus on features derived from the user's local raw dataset, such as gradients or model updates. However, they often overlook important aspects of the user's historical network traffic behavior and characteristics, such as past traffic volume and burstiness, which are strongly correlated with their local datasets~\cite{traffic-corr},~\cite{server_traffic}.

In this work, we propose a weighted attribute-based client selection strategy that identifies a more homogeneous set of users by leveraging features derived from both local data and historical network behavior. In each communication round, a composite score is computed using user-specific attributes, including historical traffic patterns, channel quality, computational capability, and prior learning contribution. Users with higher scores are more likely to exhibit similar data distributions and are thus preferred for FL training. The rationale this approach is twofold. First, traffic patterns serve as indicators of data similarity, enabling the identification of users with more aligned datasets. Second, among these users, the strategy prioritizes those with favorable channel conditions, strong computational resources, and reliable past performance, ensuring efficient training and mitigating straggler effects. By integrating these attributes into a unified weighted score, the server selects capable clients with more uniform data, improving convergence and overall FL performance.

The remainder of this paper is organized as follows. Section II introduces the network topology and traffic models, and Section III presents the proposed client selection strategy. Section IV discusses the augmented FL model, and Section V provides the simulation results. Finally, Section VI concludes the paper.

\vspace{-3mm}
\section{System Model}
\vspace{-1mm}
\subsection{Network Topology Model}
We consider an NTN architecture consisting of one HAPS located in the stratosphere and $K$ ground users.\footnote{Throughout
this work, we use the term ``client'' to refer to any ground user selected to
participate in FL training.} The HAPS acts as an FL server, providing wide
coverage and coordinating communication with geographically distributed users
within its footprint. User locations are assumed uniformly distributed. Each
user stores local data and has computational resources to perform local model
updates. As an aerial base station, the HAPS has access to historical
network-side user behavior, including traffic statistics (traffic volume
and burstiness) inferred from traffic traces, while the local data
distribution of each user remains private and is not exchanged.

\vspace{-3mm}
\subsection{Network Traffic Features Modeling}
We model uplink packet arrivals using a Cox (doubly stochastic) Poisson
process~\cite{Poisson}. Let $N_k(t)$ denote the number of packets arriving from
user $k$ at the HAPS over an interval of length $t$. Conditional on the possibly random arrival-rate process $\lambda_k(\tau)$ (packets/s),
\begin{equation}\label{eq:cox_poisson}
N_k(t)\,\big|\,\{\lambda_k(\tau)\}_{\tau\in[0,t]}
\sim \mathrm{Poisson}\!\Big(\int_{0}^{t}\lambda_k(\tau)\,d\tau\Big),
\; k\in\{1,\dots,K\},~t\ge0.
\end{equation}
In the stationary-rate case $\lambda_k(\tau)\equiv \lambda_k$,
\begin{equation}\label{eq:cox_poisson_stationary}
N_k(t)\,\big|\,\lambda_k \sim \mathrm{Poisson}(\lambda_k t).
\end{equation}

\noindent
The arrival-rate $\lambda_k$ depends on the instantaneous channel realization
$h_k$ and the packet size (in bits), denoted by $L_k$.

Given packet size $L_k$, the maximum arrival rate (packets/s) is
\begin{equation}\label{eq:max_arrival}
\lambda_{\max,k} \;=\; \frac{R_k}{L_k},
\end{equation}
where the instantaneous achievable rate (bits/s) is
\begin{equation}\label{eq:shannon_rate}
R_k \;=\; b_k \log_2\!\Bigl(1 + \frac{p_k |h_k|^2}{N_0 b_k}\Bigr),
\end{equation}
with allocated bandwidth $b_k$, transmit power $p_k$, noise power spectral density $N_0$, and
channel coefficient $h_k$.

Let $P_{\text{loss},k}$ denote the packet loss probability due to bit errors.
Then
\begin{equation}\label{eq:eff_arrival}
\lambda_k \;=\; \lambda_{\max,k}\,(1-P_{\text{loss},k})
\;=\; \frac{R_k}{L_k}\,(1-P_{\text{loss},k}).
\end{equation}
Assuming independent bit errors, for a packet of size $L_k$ bits,
\begin{equation}\label{eq:ploss_exact}
P_{\text{loss},k} \;=\; 1-(1-P_{b,k})^{L_k},
\end{equation}
where $P_{b,k}$ is the bit error rate (BER). For $M$-QAM, a common approximation is
$
P_{b,k} \approx \frac{3}{2\sqrt{M}}\,
Q\Bigl(\sqrt{\frac{3m/M - 1}{2}\,\frac{E_b}{N_0}}\Bigr),
$
where $M$ is the constellation size, $m = \log_2(M)$ is the bits per symbol, $E_b$ is the energy per bit, and $Q(\cdot)$ is the $Q$-function. Using $(1-P_b)^{L}\approx 1 + L\ln(1-P_b)$,
\begin{equation}\label{eq:lambda_approx}
\lambda_k \approx R_k\Bigl(1/L_k + \ln(1-P_{b,k})\Bigr).
\end{equation}

\noindent
Since $|h_k|^2$ and $L_k$ are random, the mean arrival rate can be written as
\begin{equation}\label{eq:gen_cox}
\mathbb{E}[\lambda_k] \;=\; \iint \lambda_k\; f_{|h_k|^2}(x)\, f_{L_k}(\ell)\,dx\,d\ell.
\end{equation}
Under Rician fading, $|h_k|^2$ follows a noncentral chi-square distribution
with 2-degree of freedom~\cite{HAPS1}. In high-SNR LoS-dominant regimes, one may
approximate $|h_k|^2$ by its mean $\Omega_k$ as follows:
\begin{equation}\label{eq:rician_mean}
\Omega_k \;=\; |d_k|^2 + 2\Psi_k^2,
\end{equation}
with LoS power $|d_k|^2$ and NLoS power $\Psi_k^2$.

Empirical studies often model packet sizes as log-normal~\cite{packet-dist}.
If $L_k\sim\mathrm{LogNormal}(\mu_k,\sigma_k^2)$, then $\mathbb{E}[1/L_k]=e^{-\mu_k+\sigma_k^2/2}$,
and from \eqref{eq:lambda_approx},
\vspace{-2mm}
\begin{align}
\mathbb{E}[\lambda_k]
&\approx
R_k\Bigl(e^{-\mu_k+\frac{\sigma_k^2}{2}} + \ln(1-P_{b,k})\Bigr), \label{eq:lambda_mean}\\
\mathrm{Var}[\lambda_k]
&\approx
R_k^2\,(e^{\sigma_k^2}-1)\,e^{-2\mu_k+\sigma_k^2}. \label{eq:lambda_var}
\end{align}

\subsubsection{Traffic Volume}
Over a window $T$ where $\mathbb{E}[\lambda_k]$ is approximately stationary,
traffic volume is
\begin{equation}\label{eq:ENk}
V_k \triangleq \mathbb{E}[N_k(T)] \;=\; \int_0^T \mathbb{E}[\lambda_k]\,dt
\;=\; T\,\mathbb{E}[\lambda_k].
\end{equation}
Though real-world network traffic may vary over time, for sufficiently short intervals, the rate can be treated as constant~\cite{def1}, making the analysis tractable.
\subsubsection{Traffic Burstiness}
We quantify traffic burstiness by the squared coefficient of variation of $\lambda_k$ as
\begin{equation}\label{eq:burstiness}
B_k \triangleq \mathrm{Var}[\lambda_k]/(\mathbb{E}[\lambda_k])^2,
\end{equation}
indicating how aggressively traffic fluctuates relative to its average. Traffic maintaining a steady rate is considered non-bursty, whereas significant peaks and valleys are bursty.
\vspace{-4mm}
%%%%%%%%%%%%%%%%%%%%%%%%%%%%%%%%%%%%%%%%%%%%%%%%%%%%%%%%%%%%%%%%%%%%%%%%%%%%%%
\subsection{Non-IIDness--Traffic Correlation}\label{subsec:niid-traffic}

In mobile services, the same user context (usage patterns, mobility, service
demand) jointly shapes (i) the local data-generation process and (ii) network
traffic behavior. The FL server does not observe the local data distribution
$P_k$ (kept private), but the HAPS can estimate historical traffic features
from traces. We therefore derive a traffic-based proxy for the latent
non-IIDness and provide a theoretical guarantee that selecting clients based on
this proxy minimizes the expected non-IIDness of the selected set.

Let $P_k$ denote user $k$'s local data distribution and
$\bar P=\sum_{j=1}^{K}\pi_j P_j$ as the global mixture distribution, where $\pi_j\ge0$ and $\sum_j\pi_j=1$. Statistical heterogeneity is quantified as
\begin{equation}\label{eq:nu_def}
\nu_k \triangleq D_{\mathrm{JS}}(P_k\|\bar P),
\end{equation}
where $D_{\mathrm{JS}}(\cdot\|\cdot)$ is the Jensen--Shannon divergence. Note
that $\nu_k$ is not directly available to the server. In addition,
let $\widehat V_k$ and $\widehat B_k$ denote HAPS-side estimates of $V_k$ and
$B_k$ obtained from historical traffic traces over an observation window $T$.

To relate traffic to heterogeneity, introduce nonnegative latent behavioral
descriptors: intensity $I_k\ge0$ and variability $Q_k\ge0$, capturing long-term
service demand and burstiness tendency, respectively. We adopt the factor model
\begin{align}
\widehat V_k &= T\bigl(a_I I_k + \xi_k\bigr), \qquad a_I>0, \label{eq:V_model}\\
\widehat B_k &= a_Q Q_k + \upsilon_k, \qquad a_Q>0, \label{eq:B_model}
\end{align}
where $\xi_k$ and $\upsilon_k$ are zero-mean estimation residuals uncorrelated
with $(I_k,Q_k)$. To connect traffic to data heterogeneity, we use the affine
approximation
\vspace{-1mm}
\begin{equation}\label{eq:nu_model}
\nu_k \;=\; \nu_0 \;-\; c_I I_k \;+\; c_Q Q_k \;+\; r_k,
\qquad c_I,c_Q>0,
\end{equation}
where $\nu_0$ is a constant offset and $r_k$ is a zero-mean mismatch term
uncorrelated with $(I_k,Q_k)$. The sign structure in \eqref{eq:nu_model}
captures the empirically common regime where sustained activity increases data
diversity (reducing divergence to $\bar P$), while bursty behavior induces
context concentration (increasing divergence). We further assume
\begin{equation}\label{eq:IQ_uncorr}
\mathrm{Cov}(I_k,Q_k)=0.
\end{equation}

Define normalized traffic features $\hat V_k,\hat B_k\in[0,1]$, and construct the traffic uniformity
attribute
\begin{equation}\label{eq:t_def}
\hat t_k \;=\; \beta_V\,\hat V_k \;+\; \beta_B\,(1-\hat B_k),
\qquad \beta_V,\beta_B\ge0,\ \beta_V+\beta_B=1.
\end{equation}
Thus, larger $\hat t_k$ corresponds to higher sustained volume and lower
burstiness, i.e., predicted lower heterogeneity under \eqref{eq:nu_model}.

The following results provide two guarantees. First, under the coupling model
\eqref{eq:V_model}--\eqref{eq:nu_model}, traffic volume and burstiness (and thus
the traffic score $\hat t_k$) are monotone predictors of the latent non-IIDness
$\nu_k$. Second, if $\mathbb{E}[\nu_k\mid \hat t_k=t]$ is strictly decreasing in
$t$, then selecting the $M$ users with the largest $\hat t_k$ minimizes the
expected average non-IIDness among all subsets of size $M$.
\begin{thm}[Traffic score as a monotone proxy for non-IIDness]
\label{thm:traffic_proxy}
Assume \eqref{eq:V_model}--\eqref{eq:IQ_uncorr} hold and
$\mathrm{Var}(I_k)>0$, $\mathrm{Var}(Q_k)>0$. Then
$\mathrm{Corr}(\nu_k,\widehat V_k)<0$ and $\mathrm{Corr}(\nu_k,\widehat B_k)>0$.
Consequently, for any normalization that is strictly increasing in
$\widehat V_k$ and strictly increasing in $1-\widehat B_k$, the resulting traffic score $\hat t_k$ in
\eqref{eq:t_def} satisfies $\mathrm{Corr}(\nu_k,\hat t_k)<0$.
\vspace*{-2mm}
\end{thm}
\proof See Appendix~\ref{app:proof_traffic_proxy}.\\
\vspace{-5mm}
\begin{thm}[Proxy-optimal minimization of expected non-IIDness]
\label{thm:proxy_opt}
Fix a round and a subset size $M$. Assume the traffic score $\hat t_k$ is
conditionally sufficient in the sense that, for every $k$,
\begin{equation}
\mathbb{E}[\nu_k \mid \{\hat t_j\}_{j=1}^K] \;=\; \mathbb{E}[\nu_k \mid \hat t_k]
\;\triangleq\; \mu(\hat t_k),
\label{eq:cond_suff}
\end{equation}
and that $\mu(t)$ exists and is strictly decreasing in $t$.\footnote{Under \eqref{eq:V_model}--\eqref{eq:nu_model}, $\mu(t)$ is decreasing in $t$ when the mismatch term $r_k$ and normalization distortion are not dominant.}
Then, among all subsets $\mathcal{S}$ with $|\mathcal{S}|=M$, selecting the $M$
users with the largest $\hat t_k$ minimizes the conditional expected average
non-IIDness
\begin{equation}\label{eq:opt_subset}
\mathcal{S}^\star \in \arg\min_{|\mathcal{S}|=M}\ 
\mathbb{E}\Big[\frac{1}{M}\sum_{k\in\mathcal{S}}\nu_k\ \Big|\ \{\hat t_j\}_{j=1}^K\Big].
\end{equation}
\vspace*{-4mm}
\end{thm}
\proof See Appendix~\ref{app:proof_proxy_opt}.\\
\vspace{-4mm}
\begin{cor}[HAPS LoS improves proxy reliability]\label{cor:haps_gain}
Let $\widehat V_k$ and $\widehat B_k$ be estimated from traffic traces over a
window of length $T$ using standard sample-mean and sample-variance estimators.
Under LoS-dominant HAPS links, reduced packet loss and more stable received
power increase observation reliability, which typically reduces the estimation
error variances of $\xi_k$ and $\upsilon_k$ as $T$ grows. Consequently, the
traffic proxy $\hat t_k$ becomes more reliable, strengthening the practical
alignment between $\hat t_k$ and the latent non-IIDness $\nu_k$.
\end{cor}

\vspace{-3mm}
%%%%%%%%%%%%%%%%%%%%%%%%%%%%%%%%%%%%%%%%%%%%%%%%%%%%%%%%%%%%%%%%%%%%%%%%%%%%%%
\section{Proposed Client Selection Strategy}
\label{sec:advanced_selection}

We propose a weighted attributes-based client selection strategy that assigns
each user a composite score based on four attributes: traffic uniformity,
channel quality, computational capability, and dynamic learning performance.

%\vspace{-1mm}
\subsection{Selection Criteria}\label{ssec:client_selection}

\paragraph{Historical Traffic Pattern (uniformity proxy)}
The HAPS constructs the traffic uniformity attribute $\hat t_k\in[0,1]$ as in
\eqref{eq:t_def} using normalized estimates $\hat V_k$ and $\hat B_k$. By
Theorems~\ref{thm:traffic_proxy}--\ref{thm:proxy_opt}, ranking users by $\hat t_k$
is theoretically justified as minimizing the conditional expected
non-IIDness of the selected set for a fixed subset size, despite the fact that
the local distribution $P_k$ is private and not shared.

\paragraph{Channel Quality}
Channel quality is represented by a normalized metric $\hat r_k\in[0,1]$ inferred
from the available SNR, e.g., $\frac{p_k|h_k|^2}{N_0b_k}$. Larger $\hat r_k$
implies better throughput and lower communication latency, and is incorporated
into the score to favor reliable links.

\paragraph{Computational Capability}
Let $f_k$ (CPU cycles/s) denote the computing capability of client $k$, which can
be estimated at the HAPS~\cite{HAPS_FL2}. Let $C_k$ be the CPU cycles required to
process one sample. With $E$ local epochs, $J_k$ samples, and local computation
time $l_k$,
\begin{equation}\label{eq:fk_def}
f_k \;=\; E\,C_k\,J_k/l_k, \quad \forall k.
\end{equation}
We normalize $f_k$ to $\hat f_k\in[0,1]$ and favor larger $\hat f_k$ to reduce
straggling.

\paragraph{Prior-round Learning Performance}
The HAPS tracks each client's historical learning contribution via the EMA score
$\hat m_k$.

\begin{lemm}\label{lemm:score}
Let $\Delta_k^{(n)}\in[0,1]$ denote the normalized loss reduction reported by
client $k$ after local training round $n$, and choose $\zeta\in(0,1)$. Define
\begin{equation}\label{eq:ema-main}
\hat m_k^{(n+1)} \;=\; \zeta\,\hat m_k^{(n)} + (1-\zeta)\,\Delta_k^{(n)}.
\end{equation}
Then: (i) $\hat m_k^{(n+1)}$ uniquely minimizes the exponentially weighted
least-squares cost
$J_{n+1}(\theta)=\sum_{t=0}^{n}\zeta^{t}(\theta-\Delta_k^{(n-t)})^{2}$;
(ii) $\hat m_k^{(n+1)}=(1-\zeta)\sum_{t=0}^{n}\zeta^{t}\Delta_k^{(n-t)}
+\zeta^{n+1}\hat m_k^{(0)}$; (iii) if $0\le \hat m_k^{(0)}\le1$, then
$0\le \hat m_k^{(n)}\le1$ for all $n$; (iv) if $\{\Delta_k^{(n)}\}$ is
mean-stationary with mean $\mu_k'$, then $\mathbb{E}[\hat m_k^{(n)}]\to\mu_k'$.
\end{lemm}
\proof See Appendix~\ref{app:score-proof}.\\
\noindent
Smaller $\zeta$ emphasizes recent rounds, while larger $\zeta$ rewards sustained
performance.

%%%%%%%%%%%%%%%%%%%%%%%%%%%%
%%%%%%%%%%%%%%%%%%%%%%%%%%%%
\begin{algorithm}[!t]
\caption{Proposed Client Selection Strategy (Round $n$)}
\label{alg:ClientSel}
{\footnotesize
\setlength{\algorithmicindent}{0.8em}
\begin{algorithmic}[1]
\REQUIRE Total users $K$; target subset size $M$; weights
$\boldsymbol{\varepsilon}^{(n)}=\{\varepsilon_t,\varepsilon_r,\varepsilon_f,\varepsilon_m\}$ with
$\varepsilon_i\ge0$, $\sum_i\varepsilon_i=1$; traffic window length $T$.
\ENSURE Selected client set $\mathcal{S}^{(n)}$.
\FOR{$k=1$ \TO $K$}
    \STATE Estimate (or update) traffic volume $\widehat V_k$ and burstiness $\widehat B_k$ from traces over window $T$.
    \STATE Normalize to $\hat V_k,\hat B_k\in[0,1]$ and compute traffic uniformity $\hat t_k$ via \eqref{eq:t_def}.
    \STATE Compute normalized channel/compute attributes $\hat r_k,\hat f_k\in[0,1]$ and retrieve $\hat m_k^{(n)}\in[0,1]$.
    \STATE Compute $\mathrm{score}_k^{(n)}=\varepsilon_t^{(n)}\hat t_k+\varepsilon_r^{(n)}\hat r_k+\varepsilon_f^{(n)}\hat f_k+\varepsilon_m^{(n)}\hat m_k^{(n)}$.
\ENDFOR
\STATE $\mathcal{S}^{(n)} \leftarrow \textbf{Top-}M$ users with the largest $\mathrm{score}_k^{(n)}$.
\STATE \textbf{Return} $\mathcal{S}^{(n)}$.
\end{algorithmic}
}
\end{algorithm}

%\vspace{-2mm}
\subsection{Composite Score Formulation}

After normalizing $\hat t_k,\hat r_k,\hat f_k,\hat m_k\in[0,1]$, the HAPS assigns
the composite score
\begin{equation}\label{eq:score}
\mathrm{score}_k^{(n)} \;=\; \varepsilon_t^{(n)} \hat t_k^{(n)} +
\varepsilon_r^{(n)} \hat r_k^{(n)} + \varepsilon_f^{(n)} \hat f_k^{(n)} +
\varepsilon_m^{(n)} \hat m_k^{(n)},
\end{equation}
where $\varepsilon_i^{(n)}\ge0$ and $\sum_{i\in\{t,r,f,m\}}\varepsilon_i^{(n)}=1$,
so $\mathrm{score}_k^{(n)}\in[0,1]$.

The term $\hat t_k$ is a uniformity proxy with theoretical guarantees:
Theorem~\ref{thm:proxy_opt} shows that, for a fixed subset size, selecting users
with the largest $\hat t_k$ minimizes the conditional expected non-IIDness
of the selected set. The remaining terms $\hat r_k,\hat f_k,\hat m_k$ capture
communication/computation/learning efficiency; thus \eqref{eq:score} is a linear
scalarization that trades non-IID reduction with efficiency.

Moreover, in round $n$, define the feasible client set
\begin{equation}\label{eq:feasible_set}
\mathcal{F}^{(n)} \triangleq \{k:\ \hat r_k^{(n)}\ge r_{\min},\ \hat f_k^{(n)}\ge f_{\min},\
\hat m_k^{(n)}\ge m_{\min}\}.
\end{equation}
Selecting the top $M$ users by $\hat t_k^{(n)}$ within $\mathcal{F}^{(n)}$ yields
the minimum conditional expected non-IIDness among all $M$-subsets of
$\mathcal{F}^{(n)}$ (by Theorem~\ref{thm:proxy_opt}), while satisfying minimum
efficiency requirements.

\subsubsection{Adaptive weight update (heuristic)}
Beyond the equal-importance baseline
$\varepsilon_t=\varepsilon_r=\varepsilon_f=\varepsilon_m=\tfrac{1}{4}$, we
adapt weights online using a short memory window of length $H$. Define the
round-level summary of attribute $i\in\{t,r,f,m\}$ as
\begin{equation}\label{eq:x_in_def}
x_i^{(n)} \triangleq \frac{1}{|\mathcal{S}^{(n)}|}\sum_{k\in\mathcal{S}^{(n)}} \hat i_k^{(n)},
\end{equation}
and the normalized global validation-loss drop as
\begin{equation}\label{eq:y_n_def}
y^{(n)} \triangleq (\mathcal{L}_\mathrm{val}^{(n-1)} - \mathcal{L}_\mathrm{val}^{(n)})/\mathcal{L}_\mathrm{val}^{(n-1)}.
\end{equation}
Over the last $H$ rounds, compute the absolute Pearson correlation
\[
\beta_i \triangleq \left|\mathrm{corr}\Big(\{x_i^{(n)}\},\{y^{(n)}\}\Big)\right|,
\]
and update
\begin{equation}\label{eq:eps_update}
\varepsilon_i^{(n)} \;=\; \frac{\beta_i}{\sum_{j\in\{t,r,f,m\}}\beta_j},\qquad
i\in\{t,r,f,m\}.
\end{equation}
This heuristic emphasizes attributes that have recently been more predictive of
global validation improvement, while preserving $\sum_i\varepsilon_i^{(n)}=1$.

\subsubsection{Client selection}
At round $n$, the selected set is
\begin{equation}\label{eq:S_select}
\mathcal{S}^{(n)} \;=\; \Big\{k:\ \mathrm{score}_k^{(n)} \ge \mathrm{score}_{\mathrm{th}}\Big\},
\end{equation}
or equivalently, one may select the top $M$ users by $\mathrm{score}_k^{(n)}$.
The feasible-set policy in \eqref{eq:feasible_set} can be used when hard minimum
efficiency constraints are required.

%%%%%%%%%%%%%%%%
%%%%%%%%%%%%%%%%%%%%%%%%%%%%
%%%%%%%%%%%%%%%%%%%%%%%%%%%
\begin{algorithm}[!t]
\caption{Augmented FL Algorithm with Weighted-Attribute Selection}
\label{alg:FL}
{\footnotesize
\setlength{\algorithmicindent}{0.8em}
\begin{algorithmic}[1]
\REQUIRE Private datasets $\{\mathcal{D}_k\}_{k=1}^K$; validation/test set $\mathcal{T}$ at HAPS;
rounds $N$; local epochs $E$; step size $\eta$; batch size $B$;
proximal parameter $\tau\ge0$; EMA factor $\zeta\in(0,1)$;
traffic window $T$; target subset size $M$; weight-memory window $H$;
initial model $\mathbf{q}^{(0)}$; initial $\{\hat m_k^{(0)}\}\subseteq[0,1]$; initial $\boldsymbol{\varepsilon}^{(0)}$.
\ENSURE Final global model $\mathbf{q}^{(N)}$ and evaluation logs.
\FOR{$n=0$ \TO $N-1$}
    \STATE $\mathcal{S}^{(n)} \leftarrow$ Algorithm~\ref{alg:ClientSel}$(K,M,\boldsymbol{\varepsilon}^{(n)},T)$. \texttt{// Client selection}
    \STATE Broadcast $\mathbf{q}^{(n)}$ to all $k\in\mathcal{S}^{(n)}$.
    \FOR{each $k\in\mathcal{S}^{(n)}$ \textbf{in parallel} } 
        \STATE Initialize $\mathbf{w}\leftarrow \mathbf{q}^{(n)}$ and run $E$ epochs SGD on $\mathcal{D}_k$ (batch $B$, step $\eta$) to minimize
        $\mathcal{L}_k(\mathbf{w};\mathcal{D}_k)+\frac{\tau}{2}\|\mathbf{w}-\mathbf{q}^{(n)}\|^2$.
        \STATE Upload $\mathbf{w}_k^{(n)}$ and report $\Delta_k^{(n)}\in[0,1]$.
    \ENDFOR
    \STATE Let $\mathcal{R}^{(n)}\subseteq\mathcal{S}^{(n)}$ be clients whose updates are received at HAPS.
    \STATE $\alpha_k^{(n)} \leftarrow \frac{|\mathcal{D}_k|}{\sum_{j\in\mathcal{R}^{(n)}}|\mathcal{D}_j|}$ for all $k\in\mathcal{R}^{(n)}$.
    \STATE $\mathbf{q}^{(n+1)} \leftarrow \sum_{k\in\mathcal{R}^{(n)}} \alpha_k^{(n)}\,\mathbf{w}_k^{(n)}$.
    \FOR{each $k\in\mathcal{R}^{(n)}$}
        \STATE $\hat m_k^{(n+1)} \leftarrow \zeta\,\hat m_k^{(n)} + (1-\zeta)\,\Delta_k^{(n)}$.
    \ENDFOR
    \STATE Evaluate $\mathbf{q}^{(n+1)}$ on $\mathcal{T}$ and record validation-loss drop $y^{(n+1)}$.
    \STATE Update $\boldsymbol{\varepsilon}^{(n+1)}$ from the last $H$ rounds using the correlation-based rule in~\eqref{eq:eps_update}.
\ENDFOR
\STATE \textbf{Return} $\mathbf{q}^{(N)}$ and recorded performance metrics.
\end{algorithmic}
}
\end{algorithm}
%%%%%%%%%%%%%%%%%%%%
%%%%%%%%%%%%%%%%%%%%
 %%%%%
\vspace{-2mm}
\section{Augmented FL Model}
\label{ssec:overall_fl_model}

We augment the standard FL procedure by integrating the proposed client
selection strategy. Global training proceeds over communication rounds
$n=0,1,\dots,N-1$. At each round $n$, the HAPS performs

\begin{enumerate}
\item \textbf{Client Selection:} Select a subset $\mathcal{S}^{(n)}\subseteq\{1,\dots,K\}$
according to the proposed weighted attribute-based selection strategy
(Algorithm~\ref{alg:ClientSel}).

\item \textbf{Model Broadcasting:} Broadcast the current global model
$\mathbf{q}^{(n)}$ to all clients in $\mathcal{S}^{(n)}$.

\item \textbf{Local Training:} Each selected client $k\in\mathcal{S}^{(n)}$
computes an updated local model by (approximately) solving
\begin{equation}\label{eq:local_obj}
\mathbf{w}_k^{(n)} \in \arg\min_{\mathbf{w}}
\Bigl\{\mathcal{L}_k(\mathbf{w};\mathcal{D}_k)
+ \frac{\tau}{2}\|\mathbf{w}-\mathbf{q}^{(n)}\|^2\Bigr\},
\end{equation}
where $\mathcal{L}_k(\mathbf{w};\mathcal{D}_k)$ is the empirical cross-entropy
loss over the local dataset $\mathcal{D}_k$, and $\tau\ge 0$ is a proximal
regularization parameter that recovers FedAvg when $\tau=0$ and yields a
FedProx-style update when $\tau>0$.

\item \textbf{Aggregation:} The HAPS aggregates the received local models to form
the next global model:
\begin{equation}\label{eq:agg}
\mathbf{q}^{(n+1)} \;=\; \sum_{k\in\mathcal{S}^{(n)}} \alpha_k^{(n)}\,\mathbf{w}_k^{(n)},
\end{equation}
with weights satisfying $\alpha_k^{(n)}\ge 0$ and $\sum_{k\in\mathcal{S}^{(n)}}\alpha_k^{(n)}=1$.
A common choice is data-size weighting,
$\alpha_k^{(n)}=\frac{|\mathcal{D}_k|}{\sum_{j\in\mathcal{S}^{(n)}}|\mathcal{D}_j|}$.
\end{enumerate}

The above steps repeat until reaching a target accuracy or the maximum number
of rounds $N$. The overall flow is summarized in Algorithm~\ref{alg:FL}.

\vspace{-4mm}
%\vspace{-1mm}
\section{Simulation Results}

In this section, we evaluate the proposed weighted attribute-based client selection strategy in a
HAPS-aided NTN FL system using CIFAR-10. Unless otherwise
stated, we consider the simulation parameters summarized in TABLE~\ref{tab:simulation_parameters}.

To capture heterogeneous service contexts, we generate client datasets with
client-specific statistical heterogeneity. Each client $k$ is assigned a
Dirichlet parameter $\alpha_k$ controlling its local label mixture: smaller
$\alpha_k$ yields stronger non-IID label skew, while larger $\alpha_k$ approaches
near-IID proportions. We couple $\alpha_k$ to the server-observable traffic
proxy $\hat t_k$, so clients with more regular traffic (higher uniformity) tend
to have more diverse labels (lower non-IIDness), whereas burstier traffic tends
to produce more concentrated label mixtures. This coupling makes $\hat t_k$ an
informative proxy for latent client heterogeneity, consistent with our analysis.

We compare the proposed strategy against the following baselines, all operating
under the same $M$ selected clients per round: (i) Random: uniform random sampling of $M$ clients, (ii)
Resource: selects clients based on a composite resource score
(channel + compute), representing common resource-aware heuristics~\cite{compare}, (ii) Traffic: selects clients with highest traffic uniformity score
$\hat t_k$%(proxy-driven selection without resource awareness)
, (iv) Channel: selects clients with best instantaneous channel quality, (v) Compute: selects clients with highest compute capacity.
%The Proposed method combines (i) channel quality, (ii) compute capacity, (iii) traffic history, and (iv) prior FL learning performance (estimated from observed loss reduction / contribution quality) into a unified selection score, with adaptive weighting based on observed utility.

\begin{figure}
     \centering
     \vspace{-2mm}
     \includegraphics[width=0.8\linewidth]{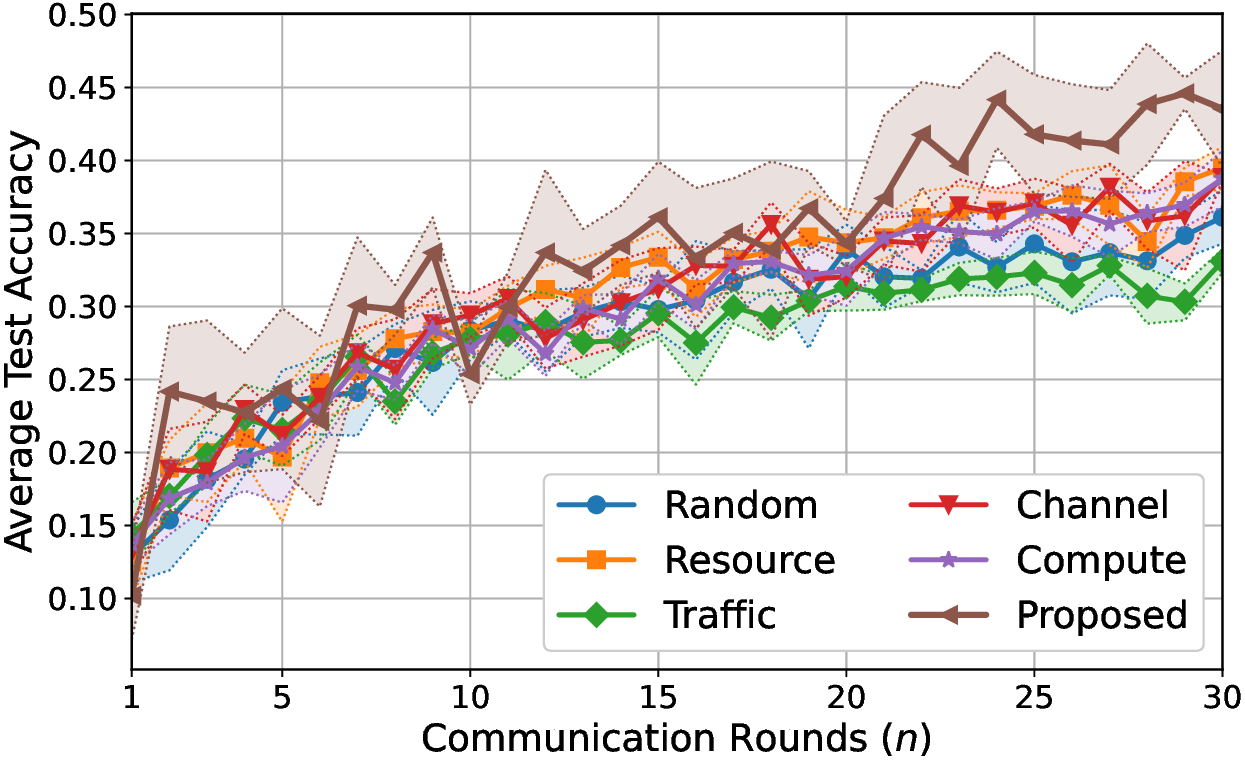}
     \caption{Average FL test accuracy over communication rounds.}
     \vspace{-4mm}
     \label{fl_acc_curve}
 \end{figure}

Fig.~\ref{fl_acc_curve} shows the average test accuracy versus communication
rounds. All methods improve with increasing rounds; however, the proposed
strategy achieves the highest accuracy trajectory and maintains a clear
advantage throughout training, with the largest separation emerging in the
mid-to-late rounds. This behavior is consistent with the design goal of
selecting clients that jointly provide (i) informative gradients (lower
effective non-IIDness via $\hat t_k$), (ii) reliable transmissions (good
channels), and (iii) sufficient local computation, while also down-weighting
clients that repeatedly yield low-quality updates.

The relative behavior of the baselines further highlights why a single-attribute
policy is insufficient in this setting. The Traffic baseline
underperforms because it ignores resource constraints: under the enforced
coupling between traffic and resources, selecting only by $\hat t_k$ can
over-select clients with weaker channels/compute, increasing dropout likelihood
and noisy updates, which degrades global learning. Conversely, the
Channel and Compute baselines improve update reliability but
do not control statistical heterogeneity, which can slow convergence under
non-IID data. The Resource baseline partially mitigates feasibility by
favoring strong channel/compute clients, yet it still lacks an explicit
mechanism to exploit heterogeneity-relevant traffic history and prior measured
contribution quality.

\begin{figure}
     \centering
     \includegraphics[width=0.801\linewidth]{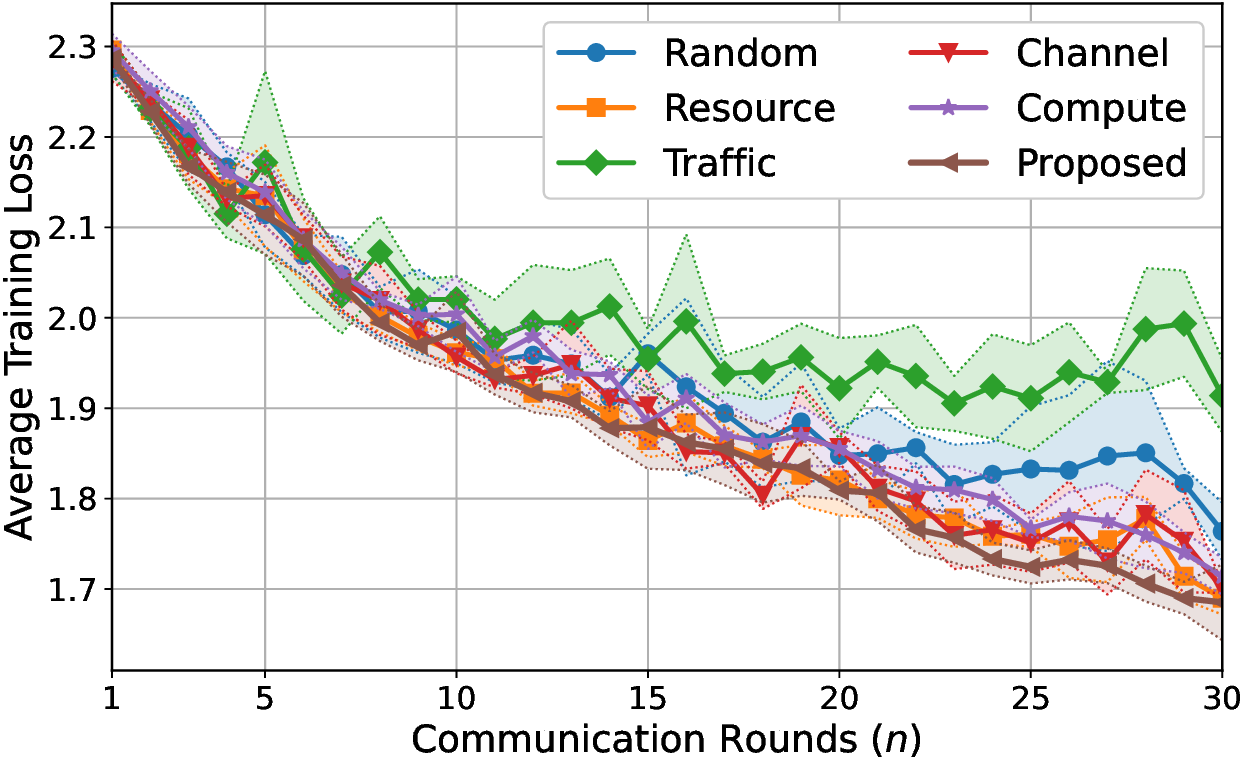}
     \caption{Average FL training loss over communication rounds.}
     \vspace{-4mm}
     \label{fl_loss_curve}
 \end{figure}

Fig.~\ref{fl_loss_curve} reports the average training/validation loss over
rounds. The proposed method yields the fastest and most consistent loss
reduction, matching the observed accuracy gains. Notably, the Traffic
baseline exhibits significantly higher loss, which aligns with its poorer
accuracy and indicates that resource-blind selection can harm optimization
progress when communication/computation bottlenecks are present. The Channel,
Compute, and Resource baselines reduce loss more effectively
than Traffic, but remain consistently above the proposed method, reflecting the
benefit of jointly optimizing selection across the four attributes.

\begin{figure}
     \centering
     \includegraphics[width=0.9\linewidth]{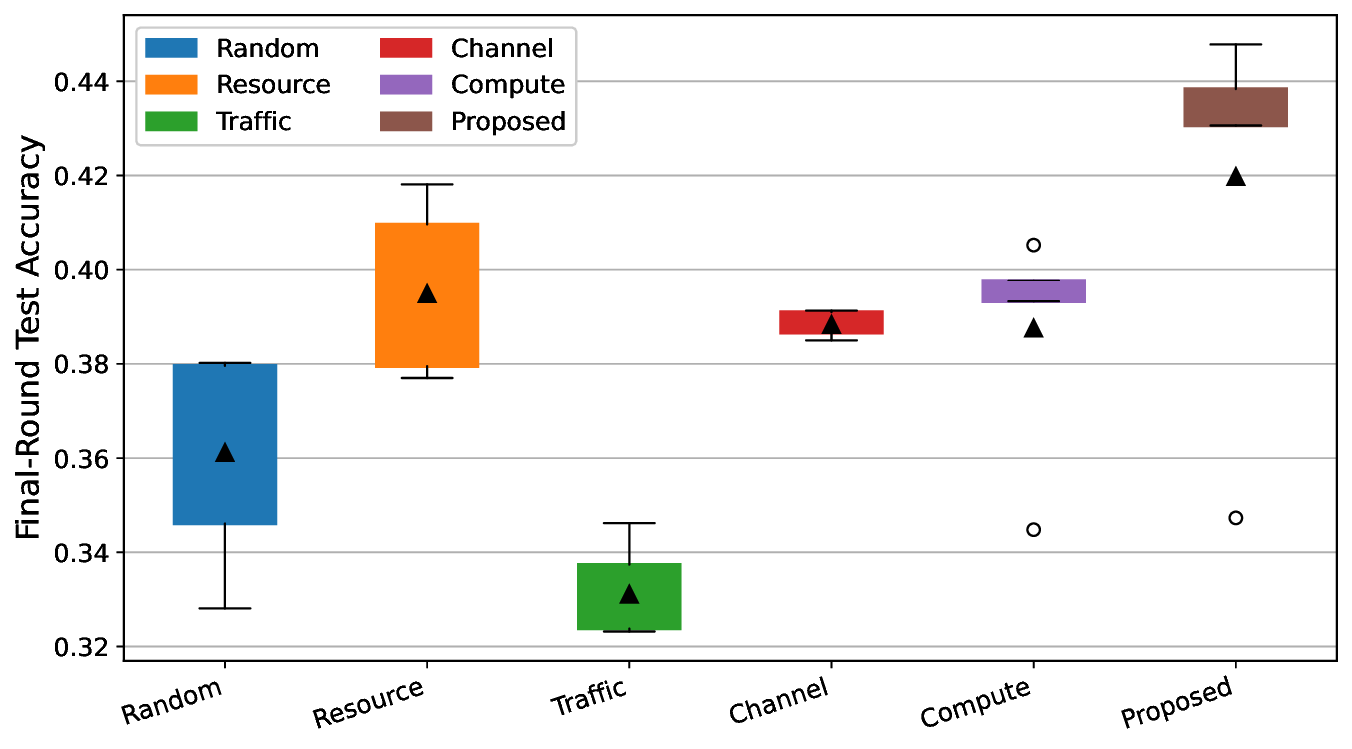}
     \caption{Final-round test accuracy distribution.}
     \label{box_final_acc}
 \end{figure}

To summarize end performance and variability, Fig.~\ref{box_final_acc}
shows the distribution of final-round test accuracy across the five seeds. The
proposed strategy achieves the best median and the highest upper range while
maintaining competitive variability, indicating that its gains are not driven
by a single favorable seed. In contrast, the baselines show lower medians and
(in some cases) wider dispersion, consistent with the greater sensitivity of
single-attribute selection to stochastic channel realizations and client
availability.

% Overall, the results support the main claim that jointly accounting for traffic behavior history, instantaneous channel quality, compute capacity, and prior observed contribution quality yields more reliable client subsets and improves learning accuracy and final generalization compared to commonly used baselines.

\vspace{-1mm}
%%%%%%%%%%%%%
%%%%%%%%%%%%%
\begin{table}[!t]
\centering
\caption{Simulation Parameters}
\vspace{-2mm}
\label{tab:simulation_parameters}
{\footnotesize
\setlength{\tabcolsep}{3pt}
\renewcommand{\arraystretch}{0.92}
\begin{tabular}{ll}
\hline
\textbf{Parameter} & \textbf{Value} \\
\hline
Dataset & CIFAR-10 \\
%Train/Test samples & 50,000 / 10,000 \\
Non-IID partitioning & Client-dependent Dirichlet label-skew \\
Dirichlet concentration $\alpha_k$ & $\alpha_k\!\in[\alpha_{\min},\alpha_{\max}]$ coupled to %traffic attribute
$\hat t_k$ \\
$\alpha_{\min}, \alpha_{\max}$ & $0.01,\;80$ \\
Samples per client & $N_k = 50{,}000/K$ \\
Total number of clients $K$ & 200 \\
Participation rate $p$ & 0.40 \\
Selected clients/round $M$ & $\lfloor pK \rfloor$ \\
Communication rounds $N$ & 30 \\
Local epochs $E$ & 4 \\
Batch size $B$ & 32 \\
Learning rate $\eta$ & $10^{-3}$ \\
FedProx coefficient $\tau$ & 0.1 \\%(FedAvg-style update) \\
Traffic normalization window $T$ & 1 (normalized) \\
Traffic attribute weights & $\beta_V=\beta_B=0.5$ \\
%Channel model & Rician fading \\
Rician $K$-factor & 10 \\
Historical performance memory $\zeta$ & 0.75 \\
Weight adaptation window $H$ & 7 rounds \\
Seeds & $\{42,43,44,45,46,47\}$ (6 runs) \\
%Client-selection policies & Random, Resource, Traffic, Channel, Compute, Proposed \\
\hline
\end{tabular}
}
\end{table}

%%%%%%%%
 %%%%%%%%
 \vspace{-1mm}
 %%%%%%%
  %%%%%%%
 %\vspace{-1mm}
 
\section{Conclusion}
We addressed non-IID data in large-scale HAPS-aided federated learning by
proposing a weighted attribute-based client selection strategy that combines
historical traffic behavior, channel quality, computational capability, and
prior-round learning contribution into a composite score. By prioritizing
clients that are expected to yield more compatible data and more reliable
updates, the proposed method reduces effective heterogeneity and improves test
accuracy, training loss, and convergence compared with random, resource-only,
and single-attribute baselines.
%%%%%%%%%%
%%%%%%%%%%
%%%%%%%%%%
\appendices
\renewcommand{\theequation}{\thesection.\arabic{equation}}

%%%%%%%%%%%%%%%%%%%%%%%%%%%%%%%%%%%%%%%%%%%%%%%%%%%%
\section{Proof of Theorem~\ref{thm:traffic_proxy}}
\label{app:proof_traffic_proxy}
\setcounter{equation}{0}
\vspace{-1mm}
\begin{proof}
From \eqref{eq:V_model} and \eqref{eq:nu_model},
$\widehat V_k=T(a_I I_k+\xi_k)$ and $\nu_k=\nu_0-c_I I_k+c_Q Q_k+r_k$.
Assume $\mathrm{Cov}(I_k,Q_k)=0$ and that the residuals $\xi_k,\upsilon_k,r_k$
are zero-mean, uncorrelated with $(I_k,Q_k)$, and mutually uncorrelated, i.e.,
$\mathrm{Cov}(r_k,\xi_k)=\mathrm{Cov}(r_k,\upsilon_k)=0$.
Then
{\small
\begin{align}
\mathrm{Cov}(\nu_k,\widehat V_k)
&=T\,\mathrm{Cov}(-c_I I_k+c_Q Q_k+r_k,\ a_I I_k+\xi_k)\nonumber\\
&=-T\,c_I a_I\,\mathrm{Var}(I_k)\;<\;0,
\end{align}
}
hence $\mathrm{Corr}(\nu_k,\widehat V_k)<0$ since $\mathrm{Var}(\nu_k)>0$ and
$\mathrm{Var}(\widehat V_k)>0$. Similarly, from \eqref{eq:B_model},
$\widehat B_k=a_Q Q_k+\upsilon_k$, and thus
{\small
\begin{align}
\mathrm{Cov}(\nu_k,\widehat B_k)
&=\mathrm{Cov}(-c_I I_k+c_Q Q_k+r_k,\ a_Q Q_k+\upsilon_k)\nonumber\\
&=c_Q a_Q\,\mathrm{Var}(Q_k)\;>\;0,
\end{align}
}
so $\mathrm{Corr}(\nu_k,\widehat B_k)>0$.

Let $\hat V_k=g_V(\widehat V_k)$ and $\hat B_k=g_B(\widehat B_k)$ be any
normalizations that are strictly increasing in their arguments. Then
$\mathrm{Corr}(\nu_k,\hat V_k)<0$ and $\mathrm{Corr}(\nu_k,\hat B_k)>0$.
Since $\hat t_k=\beta_V\hat V_k+\beta_B(1-\hat B_k)$ is strictly increasing in
$\hat V_k$ and strictly increasing in $1-\hat B_k$ for $\beta_V,\beta_B\ge0$,
$\beta_V+\beta_B=1$, it follows that $\mathrm{Corr}(\nu_k,\hat t_k)<0$.
\end{proof}
\vspace{-6mm}
%%%%%%%%%%%%%%%%%%%%%%%%%%%%%%%%%%%%%%%%%%%%%%%%%%%%
\section{Proof of Theorem~\ref{thm:proxy_opt}}
\label{app:proof_proxy_opt}
\setcounter{equation}{0}

\begin{proof}
Conditioning on $\{\hat t_j\}_{j=1}^K$ and using linearity of expectation,
{\small
\begin{align}
 \mathbb{E}\!\Big[\frac{1}{M}\sum_{k\in\mathcal{S}}\nu_k\ \Big|\ \{\hat t_j\}\Big]
=\frac{1}{M}\sum_{k\in\mathcal{S}} \mathbb{E}[\nu_k\mid \{\hat t_j\}]. 
\end{align}
}
By the conditional sufficiency assumption \eqref{eq:cond_suff},
$\mathbb{E}[\nu_k\mid \{\hat t_j\}]=\mu(\hat t_k)$, hence
\vspace{-2mm}
{\small
\begin{align}
\mathbb{E}\!\Big[\frac{1}{M}\sum_{k\in\mathcal{S}}\nu_k\ \Big|\ \{\hat t_j\}\Big]
=\frac{1}{M}\sum_{k\in\mathcal{S}}\mu(\hat t_k).    
\end{align}
}
Since $\mu(\cdot)$ is strictly decreasing, this sum is minimized by selecting
the $M$ indices with the largest $\hat t_k$.
\end{proof}
\vspace{-4mm}
%%%%%%%%%%%%%%%%%%%%%%%%%%%%%%%%%%%%%%%%%%%%%%%%%%%%
\section{Proof of Lemma~\ref{lemm:score}}
\label{app:score-proof}
\setcounter{equation}{0}
\vspace{-1mm}
\begin{proof}
\textbf{(i) Optimality.}
Define the exponentially weighted least-squares cost with an initial prior as
\vspace{-1mm}
{\small
\begin{equation}
\hspace{-2mm} J_{n+1}(\theta)
\;=\;
\zeta^{\,n+1}\bigl(\theta-\hat m_k^{(0)}\bigr)^2
\;+\;
(1-\zeta)\sum_{t=0}^{n}\zeta^{t}\bigl(\theta-\Delta_k^{(n-t)}\bigr)^2,
\end{equation}
}
where $0<\zeta<1$. Setting $\frac{\partial J_{n+1}}{\partial\theta}=0$ yields
\vspace{-1mm}
{\small
\begin{align}
% 0
% &=\zeta^{\,n+1}(\theta-\hat m_k^{(0)})
% +(1-\zeta)\sum_{t=0}^{n}\zeta^{t}(\theta-\Delta_k^{(n-t)}) \nonumber\\
% \Rightarrow\quad
\theta^\star
&=\zeta^{\,n+1}\hat m_k^{(0)}
+(1-\zeta)\sum_{t=0}^{n}\zeta^{t}\Delta_k^{(n-t)},
\end{align}
}
because $\zeta^{\,n+1}+(1-\zeta)\sum_{t=0}^{n}\zeta^t
=\zeta^{\,n+1}+(1-\zeta^{\,n+1})=1$.
Thus, $\theta^\star=\hat m_k^{(n+1)}$ and the recursion
$\hat m_k^{(n+1)}=\zeta\hat m_k^{(n)}+(1-\zeta)\Delta_k^{(n)}$ follows directly
by comparing $\theta^\star$ at rounds $n$ and $n+1$.

%\smallskip
\noindent
\textbf{(ii) Closed-form expansion.}
Unrolling the recursion gives
{\small
\begin{equation}
\hat m_k^{(n+1)}
=
(1-\zeta)\sum_{t=0}^{n}\zeta^{t}\Delta_k^{(n-t)}
+\zeta^{\,n+1}\hat m_k^{(0)}.
\end{equation}
}
%\smallskip
\noindent
\textbf{(iii) Boundedness.}
If $0\le \hat m_k^{(0)}\le1$ and $\Delta_k^{(t)}\in[0,1]$, then the closed-form
expression is a convex combination of numbers in $[0,1]$, hence
$0\le\hat m_k^{(n)}\le1$ for all $n$.

%\smallskip
\noindent
\textbf{(iv) Convergence.}
Let $E_n=\mathbb{E}[\hat m_k^{(n)}]$ and assume $\mathbb{E}[\Delta_k^{(n)}]=\mu_k'$.
Taking expectations in the recursion gives
$E_{n+1}=\zeta E_n+(1-\zeta)\mu_k'$ with solution
\vspace{-1mm}
{\small
\begin{equation}
E_n=\mu_k'(1-\zeta^{\,n})+\zeta^{\,n}\hat m_k^{(0)} \xrightarrow[n\to\infty]{}\mu_k',
\end{equation}
}
since $0<\zeta<1$.
\end{proof}
%%%%%%%%%%%%%%%%%%%%

\vspace{-7mm}
\bibliographystyle{IEEEtran}

\end{document}